\RequirePackage{ifpdf}
\ifpdf 
\documentclass[pdftex]{sigma}
\else
\documentclass{sigma}
\fi

\newcommand{\comp}{\leftrightarrow}
\newcommand{\proj}{\mbox{\rm pr}}

\newcommand{\ord}{\operatorname{ord}}

\begin{document}

\allowdisplaybreaks

\renewcommand{\thefootnote}{$\star$}

\renewcommand{\PaperNumber}{001}

\FirstPageHeading

\ShortArticleName{Archimedean Atomic Lattice Ef\/fect Algebras with Complete Lattice of Sharp Elements}

\ArticleName{Archimedean Atomic Lattice Ef\/fect Algebras \\ with Complete Lattice of Sharp Elements\footnote{This paper is a
contribution to the Proceedings of the 5-th Microconference
``Analytic and Algebraic Me\-thods~V''. The full collection is
available at
\href{http://www.emis.de/journals/SIGMA/Prague2009.html}{http://www.emis.de/journals/SIGMA/Prague2009.html}}}

\Author{Zdenka RIE\v{C}ANOV\'A}

\AuthorNameForHeading{Z. Rie\v{c}anov\'a}

\Address{Department of Mathematics,
Faculty of Electrical Engineering and Information Technology,\\
Slovak University of Technology, Ilkovi\v{c}ova~3,
SK-812~19~Bratislava, Slovak Republic}

\Email{\href{mailto:zdenka.riecanova@stuba.sk}{zdenka.riecanova@stuba.sk}}

\ArticleDates{Received September 29, 2009, in f\/inal form January 04, 2010;  Published online January 06, 2010}

\Abstract{We study Archimedean atomic lattice ef\/fect algebras
whose set of sharp elements is a complete lattice. We show
properties of centers, compatibility centers and central atoms of
such lattice ef\/fect algebras. Moreover, we prove that if such
ef\/fect algebra $E$ is separable and modular then there exists a faithful state
on $E$. Further, if an atomic lattice ef\/fect algebra is densely
embeddable into a complete
lattice ef\/fect algebra $\widehat{E}$ and the compatiblity center of $E$
is not a Boolean algebra then there exists an $(o)$-continuous
subadditive state on $E$.}


\Classification{06C15; 03G12; 81P10}

\section{Introduction, basic def\/initions and some known facts}

The classical (Kolmogorovian) probability theory
is assuming that every two events are simultaniously
measurable (as every two elements of a Boolean algebra are
mutually compatible). Thus this theory cannot explain events
occuring, e.g., in quantum physics,
as well as in economy and many other areas.

      Ef\/fect algebras generalize orthomodular lattices (including noncompatible pairs of elements) and $MV$-algebras (including unsharp elements, meaning
that $x$ and non $x$ are not disjoint). Thus ef\/fect algebras may be carriers
of probability measures when elements of these structures represent
properties, questions or events with fuzziness, uncertainty or unsharpness,
and which may be mutually non-compatible. For equivalent (in  some sense)
structures called D-posets introduced by F.~K\^opka and F.~Chovanec
we refer the reader to \cite{dvurec} and citations given there.

      Ef\/fect algebras were introduced by D.J.~Foulis and
M.K.~Bennet (see~\cite{FoBe})
for modelling unsharp measurements in a Hilbert space. In this case the
set $E(H)$ of ef\/fects is the set of all self-adjoint operators $A$ on
a Hilbert space $H$ between the null operator $0$ and the identity
operator $1$ and endowed with the partial operation $+$ def\/ined
if\/f  $A+B$ is in $E(H)$, where $+$ is the usual operator sum.
The ef\/fect algebra $E(H)$ is called a {\em standard effect algebra of
operators on Hilbert space} (or, a {\em standard Hilbert space effect algebra},
for short).

      In approach to the mathematical foundations of physics the
fundamental notions are states, observables and symmetries.
D.J.~Foulis~\cite{foulis2007}
showed how ef\/fect algebras
arise in physics and how they can be used to tie together
the observables, states  and symmetries employed in the study of physical systems.
Moreover, that due to work of G.~Ludwig~\cite{ludwig}
and subsequent book~\cite{BGL}
by  P.~Bush, P.J.~Lahti and P.~Mittelstaedt
a fourth fundamental notion, called {\em effect}, should be appended to this list.

     In spite of the fact that ef\/fect algebras are very natural algebraic structures
to be carriers of states and probability measures, in above mentioned non-classical
cases of sets of events, there are ef\/fect algebras admitting no states and hence
also no probability measures (see~\cite{ZR62}). Questions about the existence of states on ef\/fect algebras
are at present time only partially solved. They are known only some families
of ef\/fect algebras with states.

    We study Archimedean atomic lattice ef\/fect algebras with a complete lattice
of sharp elements as well as such which are densely embedable into complete
atomic lattice ef\/fect algebras. For these ef\/fect algebras we give conditions
under which states on them exist.

\begin{definition}[\cite{FoBe}]\label{def:EA}
A partial algebra $(E;\oplus,0,1)$ is called an {\em effect algebra} if
$0$, $1$ are two distinct elements and $\oplus$ is a partially
def\/ined binary operation on $E$ which satisfy the following
conditions for any $x,y,z\in E$:
\begin{description}\itemsep=0pt
\item[$(Ei)$\phantom{iii}] $x\oplus y=y\oplus x$ if $x\oplus y$ is def\/ined,
\item[$(Eii)$\phantom{ii}] $(x\oplus y)\oplus z=x\oplus(y\oplus z)$  if one
side is def\/ined,
\item[$(Eiii)$] for every $x\in E$ there exists a unique $y\in
E$ such that $x\oplus y=1$ (we put $x'=y$),
\item[$(Eiv)$\phantom{i}] if $1\oplus x$ is def\/ined then $x=0$.
\end{description}

\end{definition}

We often denote the ef\/fect algebra $(E;\oplus,0,1)$ brief\/ly by
$E$. On every ef\/fect algebra $E$  the partial order
$\le$  and a partial binary operation $\ominus$ can be
introduced as follows:
\[
x\le y  \mbox{ and }  y\ominus x=z  \mbox{ if\/f } x\oplus z
\mbox{ is def\/ined and } x\oplus z=y .
\]

If $E$ with the def\/ined partial order is a lattice (a complete
lattice) then $(E;\oplus,0,1)$ is called a {\em lattice effect
algebra} ({\em a complete lattice effect algebra}).

\begin{definition}\label{subef}
Let $E$ be an  ef\/fect algebra.
Then $Q\subseteq E$ is called a {\em sub-effect algebra} of  $E$ if
\begin{enumerate}\itemsep=0pt
\item[$(i)$] $1\in Q$,
\item[$(ii)$] if out of elements $x,y,z\in E$ with $x\oplus y=z$
two are in $Q$, then $x,y,z\in Q$.
\end{enumerate}
If $E$ is a lattice ef\/fect algebra and $Q$ is a sub-lattice and a sub-ef\/fect
algebra of $E$ then $Q$ is called a~{\em sub-lattice effect algebra} of $E$.
\end{definition}

Note that a sub-ef\/fect algebra $Q$
(sub-lattice ef\/fect algebra $Q$) of an  ef\/fect algebra $E$
(of a~lattice ef\/fect algebra $E$) with inherited operation
$\oplus$ is an  ef\/fect algebra (lattice ef\/fect algebra)
in its own right.


For an element $x$ of an ef\/fect algebra $E$ we write
$\ord(x)=\infty$ if $nx=x\oplus x\oplus\dots\oplus x$ ($n$-times)
exists for every positive integer $n$ and we write $\ord(x)=n_x$
if $n_x$ is the greatest positive integer such that $n_xx$
exists in $E$.  An ef\/fect algebra $E$ is {\em Archimedean} if
$\ord(x)<\infty$ for all $x\in E$.

A minimal nonzero element of an ef\/fect algebra  $E$
is called an {\em atom}  and $E$ is
called {\em atomic} if under every nonzero element of
$E$ there is an atom.

For a poset $P$ and its subposet $Q\subseteq P$ we denote,
for all $X\subseteq Q$, by $\bigvee_{Q} X$ the join of
the subset $X$ in the poset $Q$ whenever it exists.

We say that a f\/inite system $F=(x_k)_{k=1}^n$ of not necessarily
dif\/ferent elements of an ef\/fect algebra $(E;\oplus,0,1)$ is
{\it orthogonal} if $x_1\oplus x_2\oplus \cdots\oplus
x_n$ (written $\bigoplus\limits_{k=1}^n x_k$ or $\bigoplus F$) exists
in $E$. Here we def\/ine $x_1\oplus x_2\oplus \cdots\oplus x_n=
(x_1\oplus x_2\oplus \cdots\oplus x_{n-1})\oplus x_n$ supposing
that $\bigoplus\limits_{k=1}^{n-1}x_k$ is def\/ined and
$\bigoplus\limits_{k=1}^{n-1}x_k\le x'_n$. We also def\/ine
$\bigoplus \varnothing=0$.
An arbitrary system
$G=(x_{\kappa})_{\kappa\in H}$ of not necessarily dif\/ferent
elements of $E$ is called {orthogonal} if $\bigoplus K$
exists for every f\/inite $K\subseteq G$. We say that for an {orthogonal}
system $G=(x_{\kappa})_{\kappa\in H}$ the
element $\bigoplus G$ exists if\/f
$\bigvee\{\bigoplus K
\mid
K\subseteq G$ is f\/inite$\}$ exists in $E$ and then we put
$\bigoplus G=\bigvee\{\bigoplus K\mid K\subseteq G$ is
f\/inite$\}$. (Here we write $G_1\subseteq G$ if\/f there is
$H_1\subseteq H$ such that $G_1=(x_{\kappa})_{\kappa\in
H_1}$).

\section{The center of Archimedean atomic lattice ef\/fect algebra\\
with complete lattice of sharp elements}

The notions of a sharp element and a sharply dominating ef\/fect algebra
are due to S.~Gud\-der~\cite{gudder1,gudder2}. An element
$w$ of an ef\/fect algebra $E$ is called
{\em sharp} if $w\wedge w'=0$, and $E$
is called {\em sharply dominating} if for every $x\in E$ there
exists a smallest sharp element $w$ with
the property $x\leq w$.

The well known fact is that in every lattice ef\/fect algebra $E$
the subset $S(E)=\{w\in E\mid w\wedge w'=0\}$ is a sub-lattice
 ef\/fect algebra of $E$ being an orthomodular lattice. Moreover if
 for $D\subseteq S(E)$ the element $\bigvee_{E}D$ exists then
 $\bigvee_{E}D\in S(E)$  hence $\bigvee_{S(E)}D=\bigvee_{E}D$.
We say that $S(E)$  is a full sublattice of $E$ (see \cite{ZR57}).

Recall that elements $x, y$ of a lattice ef\/fect algebra
$(E;\oplus,0,1)$ are called {\it compatible} (written
$x\comp y$) if\/f
$x\vee y=x\oplus(y\ominus(x\wedge y))$ (see \cite{kopkabool}).
$P\subseteq E$ is a {\it set of pairwise compatible elements}
if $x\leftrightarrow y$ for all $x,y\in P$.  $M\subseteq E$ is
called a {\it block} of $E$ if\/f $M$ is a maximal subset of
pairwise compatible elements. Every block of a lattice ef\/fect
algebra $E$ is a sub-ef\/fect algebra and a~sub-lattice of $E$ and
$E$ is a union of its blocks (see \cite{ZR56}).
Lattice ef\/fect algebra with a unique block is called an
{\it $MV$-effect algebra}. Every block of a lattice ef\/fect
algebra is an $MV$-ef\/fect algebra in its own right.

An element $z$ of an ef\/fect algebra $E$ is called {\it central}
if $x=(x\wedge z)\vee (x\wedge z')$ for all $x\in E$. The {\it
center} $C(E)$ of $E$ is the set of all central elements of $E$, see
\cite{GrFoPu}. If $E$ is a lattice ef\/fect algebra then $z\in E$ is
central if\/f $z\wedge z'=0$ and $z\comp x$ for all $x\in E$, see \cite{ZR52}.
Thus in a lattice ef\/fect algebra $E$, $C(E)=B(E)\cap
S(E)$, where $B(E)=\bigcap\{M\subseteq E\mid M\text{ is a block
of }E\}$ is called a~{\it compatibility center} of $E$.
Evidently, $B(E)=\{x\in E\mid x\comp y\text{ for
all }y\in E\}$  and $B(E)$ is an $MV$-ef\/fect
algebra. Hence $C(E)$ is a Boolean algebra, see~\cite{GrFoPu}. Moreover,
$B(E)$, $C(E)$ and every block $M$ are also full sub-lattices
of a lattice ef\/fect algebra $E$. Further, $B(E)$ and $C(E)$ are sub-ef\/fect
algebras of $E$, see~\cite{GrFoPu,ZR57,ZR60}.

We are going to show that
if the set $S(E)$ of an Archimedean atomic lattice ef\/fect algebra is
a complete lattice then $E$ is isomorphic to
a subdirect product of irreducible lattice ef\/fect algebras.

\begin{definition}\label{dirprod}A direct product $\prod\{E_{\kappa}\mid\kappa\in
H\}$ of ef\/fect algebras $E_{\kappa}$ is a cartesian product
with $\oplus$, $0$, $1$ def\/ined ``coordinatewise'', i.e.,
$(a_{\kappa})_{\kappa\in H}\oplus(b_{\kappa})_{\kappa\in H}$ exists if\/f
$a_{\kappa}\oplus_{\kappa}b_{\kappa}$ is def\/ined for each $\kappa\in H$ and then
$(a_{\kappa})_{\kappa\in H}\oplus(b_{\kappa})_{\kappa\in H}=
\bigl(a_{\kappa}\oplus_{\kappa}b_{\kappa}\bigr)_{\kappa\in H}$.
Moreover, $0=(0_{\kappa})_{\kappa\in H}$, $1=(1_{\kappa})_{\kappa\in
H}$.

A \emph{subdirect product} of a family $\{E_{\kappa}\mid
\kappa\in H\}$ of lattice ef\/fect algebras is a sublattice-ef\/fect algebra $Q$
(i.e., $Q$ is simultaneously a sub-ef\/fect algebra and a
sublattice) of the direct product $\prod\{E_{\kappa}\mid
\kappa\in H\}$ such that each restriction of the natural projection
$\proj_{\kappa_i}$ to $Q$ is onto $E_{\kappa_i}$.
\end{definition}

For every central element $z$ of a lattice ef\/fect algebra $E$,
the interval $[0,z]$ with the $\oplus$ operation inherited from
$E$ and the new unity $z$ is a lattice ef\/fect algebra in its own
right.
We are going to prove a statement about decompositions of $E$ into
subdirect products of such intervals $[0,z]$ with $C([0, z])=\{0,z\}$
in the case when $E$ is an Archimedean atomic lattice ef\/fect algebra
with~$S(E)$ being a complete orthomodular lattice.

It was proved in \cite[Lemma 2.7]{PR6} that {\em in every
Archimedean atomic lattice effect algebra $E$
the subset $S(E)$ is a bifull sub-lattice of $E$},
meaning that for any $D\subseteq S(E)$ the element
$\bigvee_{S(E)} D$ exists if\/f the element $\bigvee_{E} D$ exists and
they are equal. However, as M.~Kalina proved, the center~$C(E)$ of an Archimedean atomic lattice ef\/fect algebra~$E$
need not be a bifull sub-lattice of $E$. Namely, if~$C(E)$ is an atomic Boolean algebra, then
$\bigvee_{E} \{p\in C(E) \mid p\ \mbox{atom}\ \mbox{of}\ C(E)\}$
need not exist.

\begin{definition}\label{Definition 1.3} Let $(E;\oplus_E,0_E,1_E)$
and $(F;\oplus_F,0_F,1_F)$ be ef\/fect algebras. A bijective map
$\varphi{}:E\to F$ is called an {\it isomorphism} if
\begin{enumerate}\itemsep=0pt
\item[$(i)$]
$\varphi(1_E)=1_F$,
\item[$(ii)$]
for all $a,b\in E$: $a\le_E b'$ if\/f $\varphi(a)\le_F\bigl(
\varphi(b)\bigr)'$ in such case $\varphi(a\oplus_E b)=
\varphi(a)\oplus_F\varphi(b)$.
\end{enumerate}
We write $E\cong F$. Sometimes we identify $E$ with
$F=\varphi(E)$.
If $\varphi{}:E\to F$ is an injection with properties (i) and (ii)
then $\varphi$ is called an {\it embedding\/}.
\end{definition}

\begin{lemma}\label{xdcv} Let $E$ be an Archimedean atomic lattice effect algebra. The following
conditions are equivalent:
\begin{enumerate}\itemsep=0pt
\item[$(i)$] $C(E)$ is atomic and
$\bigvee_{E}\{ p\in C(E) \mid p\ \mbox{atom}\ \mbox{of}\ C(E)\}=1$.
\item[$(ii)$] For every atom $a$ of $E$ there exists an atom
$p_a$ of $C(E)$ with $a\leq p_a$.
\item[$(iii)$] $E$ is isomorphic to a subdirect product
of the family $\{ [0, p] \mid p\ \mbox{atom}\ \mbox{of}\ C(E)\}$.
\end{enumerate}
\end{lemma}

\begin{proof} $(i)\implies (iii)$: By \cite[Theorem 3.1]{ZR71}
the condition $(i)$ implies that  $E$ is isomorphic to a~subdirect product
of the family $\{ [0, p] \mid p\ \mbox{atom}\ \mbox{of}\ C(E)\}$, since
evidently $p_1\wedge p_2=0$ for every pair of mutually dif\/ferent atoms
$p_1, p_2$ {of}\ $C(E)$.

$(iii) \implies (ii)$: By def\/inition of the
direct and subdirect products, if $a$ is an atom of $E$ then
$a\in [0, p]$ for some atom $p$ of $C(E)$.

$(ii) \implies  (i)$: Every Archimedean atomic lattice ef\/fect algebra
$E$ is a union of its atomic blocks \cite{mosna}. Let $M$ be an atomic block of
$E$ and let $A_M=\{a\in M \mid a\ \mbox{atom of}\ M\}$. Then every $a\in A_M$
is an atom of $E$. Otherwise,
for $b\in E$ with $b < a$ we have $b < a\leq x$ or $b < a\leq x'$ for all
$x\in M$, as $a\comp x$. It follows that $b\comp x$. The
maximality of blocks gives that $b\in M$, a~contradiction.

Let $a\in A_M$. As $C(E)\subseteq S(E)$,
if $a\leq z\in S(E)$ then $n_a a\leq z$. It follows that
$a\leq p_a \implies n_a a\leq p_a$ since
$p_a\in C(E)\subseteq S(E)$. Since by \cite{PR6}
$\bigvee_{E}\{n_a a\in M \mid a\ \mbox{atom of}\ M\}=1$, we obtain
that $\bigvee_{E}\{p\in C(E) \mid a\ \mbox{atom of}\ C(E)\}=1$.
\end{proof}

\begin{theorem}\label{xxx} Let $E$ be an Archimedean atomic
lattice effect algebra. Let the set $S(E)$ of sharp
elements of $E$ be a complete lattice. Then
\begin{enumerate}\itemsep=0pt
\item[$(i)$] $C(E)$ is a complete Boolean algebra and
a bifull sub-lattice of $E$.
\item[$(ii)$] $C(E)$ is atomic with
$\bigvee_{E}\{p\in C(E) \mid p\ \mbox{atom of}\ C(E)\}=1$.
\item[$(iii)$] For every atom $a$ of $E$ there exists an atom
$p$ of $C(E)$ with $a\leq p$.
\item[$(iv)$] $E$ is sharply dominating and
$E$ is isomorphic to a subdirect product of irreducible lattice effect algebras.
\end{enumerate}
\end{theorem}
\begin{proof} $(i)$: By \cite{ZR60}, $S(E)$ is a full sublattice
of $E$. By \cite[Lemma 2.7]{PR6}, in every Archimedean atomic
lattice ef\/fect algebra $E$ the set $S(E)$ is a bifull sub-lattice of $E$.
Let $D\subseteq C(E)$. Since
$S(E)$ is complete and $C(E)=S(E)\cap B(E)\subseteq S(E)$ there exists
$\bigvee_{S(E)} D=\bigvee_{E} D\in C(E)$. This proves
that $\bigvee_{C(E)} D=\bigvee_{E} D$,
which proves that $C(E)$ is a complete and bifull sub-lattice  of $E$.

$(ii)$, $(iii)$: Let $a$ be an atom of $E$ and let
$p_a=\bigwedge_{C(E)} \{ z\in C(E) \mid a\leq z\}=
\bigwedge_{E} \{ z\in C(E) \mid a\leq z\}$, using $(i)$.
Assume that there exists $c\in C(E)$, $c\not =0$ and
$c < p_a$. Then $c\in B(E)$ and hence $c\comp a$, which gives that
$a\leq c$ or $a\leq c'$. Since $a\leq c$ $\implies$ $p_a\leq c < p_a$
and $a\leq c'$  $\implies$ $ c < p_a\leq c'$, a contradiction in both
cases. Hence $c\in C(E)$ with  $0\not =c< p_a$ does not exist. It follows that
$p_a$ is an atom of $C(E)$ such that $ a\leq  p_a$. By (i),
$1=\bigvee_{C(E)}\{p\in C(E) \mid p\ \mbox{atom of}\ C(E)\}=%
\bigvee_{E}\{p\in C(E) \mid p\ \mbox{atom of}\ C(E)\}$.

$(iv)$: Since $S(E)$ is a  complete and bifull
sub-lattice of $E$, the  lattice ef\/fect algebra $E$ is
sharply dominating. Moreover, $E$ is isomorphic to a subdirect product of lattice ef\/fect algebras by Lemma~\ref{xdcv}. Let us show that,  for every \mbox{atom}~$p$ \mbox{of}~$C(E)$,
the interval $[0, p]$  is an irreducible lattice ef\/fect algebra.

Assume to the contrary that there exist
$p_0\in A_{C(E)}=\{p\in C(E) \mid p\ \mbox{atom of}\ C(E)\}$ and
$x_{p_0}\in E$ such that $x_{p_0}\in C([0, p_0])$ and $0<x_{p_0}<p_0$.
Let $x=(x_p)_{p\in A_{C(E)}}$ be such that $x_p=0$ for all $p\not = p_0$.
Then $x\wedge x'=0$. Further, let
$y\in E$ be an arbitrary. Then
$y=(y_p)_{p\in A_{C(E)}}=\bigvee_E\{y_p \mid {p\in A_{C(E)}}\},$
which gives that $x\comp y$ as $x_p\comp y_p$ for every
$p\in A_{C(E)}$ (see~\cite{ZR57}). Thus
$x\in S(E)\cap B(E)=C(E)$ which contradicts
$0<x_{p_0}<p_0$.
\end{proof}

Note that in every complete (hence in every f\/inite)
lattice ef\/fect algebra $E$ the set $S(E)$ is a complete sub-lattice of~$E$ (see~\cite{ZR57}). Let us show an example of an
Archimedean atomic $MV$-ef\/fect algebra $M$ which $S(M)$
is not a complete lattice.

\begin{example}
Simplest examples of $MV$-ef\/fect algebras are
f\/inite chains $M_k=\{0_k, a_k, 2a_k, \dots,$ $ n_{a_k} a_k = 1_k\}$,
$k=1, 2, \dots$. In this case $S(M_k)=C(M_k)=\{0_k, 1_k\}$ and
$B(M_k)=M_k$, for $k=1, 2, \dots$. Let $M$ be a direct product
of a family $\{M_k\mid  k=1, 2, \dots\}$ of $MV$-ef\/fect algebras.
Then $\oplus$, $0, 1$ on $M=\overset{\infty}{\underset{k=1}{\prod}}M_k$ are
def\/ined ``coordinatewise'' (see Def\/inition~\ref{dirprod}).

It follows that if $x=(x_k)_{k=1}^{\infty}\in M$, hence
$x_k \in M_k$ ($k=1,2,\dots$) then $x'=(x'_k)_{k=1}^{\infty}$. If
$x_k \not= 0_k$ for at most f\/inite set of $k\in\{1, 2, \dots \}$ then
$x$ is called a {\em finite element}. Clearly $x'$ is f\/inite if\/f
$x_k \not= n_{a_k} a_k$ for at most f\/inite set of $k\in\{1, 2, \dots \}$.

Let $M^{*}=\{x\in M\mid \mbox{either}\ x\ \mbox{or}\ x'\ \mbox{is f\/inite}\}$.
We can easily see that $M^{*}$ is a sub-$MV$-ef\/fect algebra (i.e.,
a sub-lattice and a sub-ef\/fect algebra of $E$). Evidently $M^{*}$ as well
as $S(M^{*})=C(M^{*})$ are not complete. Nevertheless,
$\bigvee_M\{p\in C(M^{*}) \mid p\ \mbox{atom of}\ C(M^{*})\}=1$. Here
$p\in C(M^{*})$ is an atom of $C(M^{*})$ if\/f
$p=(p_k)_{k=1}^{\infty}$ such that there exists $k_0$ with
$p_{k_0}=n_{a_{k_0}} a_{k_0}$ and $p_{k}=0$ for all $k\not= k_0$.
\end{example}

Recently, M.~Kalina showed an example of an
Archimedean atomic lattice ef\/fect algebra~$E$ with atomic $C(E)$
for which the element
$\bigvee_E\{p\in C(E) \mid p\ \mbox{atom of}\ C(E)\}$ does not
exist (see~\cite{kalina}). Thus, by Theorem~\ref{xxx}~$(ii)$, $S(E)$ is not
a complete lattice.

\section{Applications in questions about the existence of
faithful states\\ on modular lattice ef\/fect algebras}

If a lattice ef\/fect algebra $E$ is a modular lattice then $E$
is called a {\em modular lattice effect algebra}. Recall the
def\/inition of a state on ef\/fect algebras.

\begin{definition}\label{Dstate}
A map $\omega:E\to [0,1]$ is called a
{\em state on the effect algebra} $E$ if $(i)$~$\omega(1)=1$ and $(ii)$~$\omega(x\oplus
y)=\omega(x)+\omega(y)$ whenever $x\oplus y$ exists in $E$.
If, moreover, $E$ is lattice ordered then $\omega$  is
called {\em subadditive} if $\omega(x\vee y)\leq \omega(x)+\omega(y)$,
for all $x, y\in E$.
\end{definition}

Note that a state on a lattice ef\/fect algebra need not be subadditive,
while every state on an $MV$-ef\/fect algebra is subadditive.
In \cite{ZR63} it was proved that a state $\omega$ on
a lattice ef\/fect algebra $E$ is subadditive if\/f
$\omega$ is a {\em valuation}, meaning that, for any $x, y \in E$,
$x\wedge y = 0$ implies $\omega(x\vee y)= \omega(x) + \omega(y)$.

A state $\omega$ on an ef\/fect algebra $E$ is called
{\em faithful\/} if $\omega(x)>0\implies x>0$ for any
$x\in E$.

An Archimedean ef\/fect algebra is called
{\it separable\/} if every $\oplus$-orthogonal system $M$ of elements of
$E$ (meaning that for every f\/inite subset $\{x_1,x_2,\dots,x_n\}\subseteq  M$
there exists $x_1\oplus x_2\oplus \cdots \oplus x_n$) is at most countable.

\begin{theorem} Let $E$ be a separable Archimedean atomic
modular lattice effect algebra. Let the set $S(E)$ of sharp
elements of $E$ be a complete lattice. Then there exists a
faithful state on $E$.
\end{theorem}
\begin{proof} By Theorem \ref{xxx}, $C(E)$ is a complete atomic Boolean
algebra. Clearly, the orthomodular lattice $S(E)$ and the Boolean
algebra $C(E)$ are separable. Hence $C(E)$ has only countably
many atoms. It follows that there exists an $(o)$-continuous
faithful state $\omega$ on $C(E)$. Since $C(E)$ is a~Boolean algebra,
the state $\omega$ is subadditive, which is equivalent
to the fact that $\omega$ is a valuation (see \cite[Theorem~2.3]{ZR63}).
This implies the existence of
an $(o)$-continuous  faithful state  on~$S(E)$
(see \cite[Theorem~5, p.~91]{sarymsakov}). Since $S(E)$  is a complete lattice,
which is a bifull sub-lattice of $E$ (see~\cite{PR6}), the ef\/fect algebra $E$
is evidently sharply dominating. Now, using Theorem~4.3 of \cite{wujunde} and its
proof we obtain that there exists a faithful state on $E$,
as $n_a a\in S(E)$ for every atom $a$ of $E$.
\end{proof}

\section{Atomic lattice ef\/fect algebras densely embeddable\\ into
complete  lattice ef\/fect algebras}

It is well known that  any partially ordered set  $L$ can be
embedded into a complete lattice $\widehat{L}$ by an algebraic
method called {\em MacNeille completion} (or {\em completion by cuts}).
It was proved in~\cite{schmidt} that any complete lattice $K$
into which $L$ can be join-densely and meet-densely embedded is
isomorphic to its MacNeille completion $\widehat{L}$. It means that
to every element $x\in L$ there exist $M,Q\subseteq L$ such that
$x=\bigvee_{\widehat{L}}\varphi(M)=\bigwedge_{\widehat{L}}\varphi(Q)$
(here we usually  identify $L$ with $\varphi(L)$,
where $\varphi:L\to \widehat{L}$ is the embedding).

However, there are lattice ef\/fect algebras $(E;\oplus,0,1)$
for which a complete lattice ef\/fect algebra
$(\widehat{E};\widehat{\oplus}, 0,1)$
with above mentioned properties does not exist
($\oplus$ from~$E$ cannot be extended  onto~$\widehat{E}$, see~\cite{ZR53}).

In this part we study properties of central atoms of atomic
lattice ef\/fect algebras for which the MacNeille completion
$(\widehat{E};\widehat{\oplus}, 0,1)$ exists
(see \cite{ZR53} for necessary and suf\/f\/icient conditions for that).
Note that if this is the case, then $E$ is Archimedean
because every complete lattice ef\/fect algebra  is Archimedean
(\cite[Theorem 3.3]{ZR53}).

\begin{lemma}\label{mcne}
Let $(E;\oplus,0,1)$ be an atomic lattice effect
algebra and let $(\widehat E;\widehat{\oplus}, 0, 1)$ be
its MacNeille completion.
Then
\begin{enumerate}\itemsep=0pt
\item[$(i)$] $E$ and $\widehat E$ are
Archimedean atomic lattice effect
algebra with the same set of atoms $A_E$.
\item[$(ii)$] To every maximal subset
of pairwise compatible atoms $A\subseteq A_E=A_{\widehat E}$
uniquely exist an atomic block  $M$ of $E$ and an atomic block
 $\widehat M$ of $\widehat E$ such that $\widehat  M \cap E=M$ and
$\widehat M$ is the MacNeille completion of the block $M$.
\item[$(iii)$] $C(\widehat E)$ is an atomic Boolean algebra.
\item[$(iv)$] $S(\widehat E)$ is the MacNeille completion of
$S(E)$ and $S(E)=S(\widehat E)\cap E$.
\item[$(v)$] $S(E)$ is atomic iff $S(\widehat E)$ is atomic
and then their sets of all atoms coincide.
\end{enumerate}
\end{lemma}

\begin{proof} We identify $E$ with $\varphi(E)\subseteq \widehat{E}$,
where  $\varphi:E\to \widehat{E}$ is the embedding (see~\cite{schmidt}).

$(i)$: Since $E\subseteq \widehat{E}$, we have that
$A_E\subseteq A_{\widehat{E}}$, where
$A_E$ and $A_{\widehat{E}}$ are sets of atoms of $E$ and
$\widehat{E}$, respectively. Further the fact that $E$ is join-dense
in $\widehat{E}$ implies that
$x=\bigvee_{\widehat{E}}\{ y\in E \mid y\leq x\}$, for
every $x\in \widehat{E}$, which gives that
$A_{\widehat{E}}\subseteq E$ and hence
$A_E= A_{\widehat{E}}$. By \cite[Theorem 3.3]{ZR54}
$\widehat{E}$ is Archimedean, which implies that $E$
is also Archimedean.

$(ii)$: By \cite{mosna} every Archimedean atomic lattice ef\/fect
algebra~$E$ is a union of its atomic blocks~$M$ which uniquely
correspond to a maximal subset $A_M$ of pairwise compatible atoms of~$E$. Thus $A_M\subseteq M \subseteq{\widehat{M}}$ and since
$\widehat{M}$ is a complete $MV$-ef\/fect algebra with the same set
 of atoms as $M$ we obtain that
${\widehat{M}}$ is the MacNeille completion of the block $M$ by
the Schmidt characterization (see~\cite{schmidt}).

$(iii)$: Since every complete atomic lattice ef\/fect algebra
is orthocomplete, it has atomic center (see~\cite{JePu}), we obtain that
$C(\widehat{E})$ is atomic.

$(iv)$: Let $y\in S(\widehat{E})$. Then there exists a set
$\{ a_{\kappa} \mid \kappa\in H\}\subseteq A_{\widehat{E}}=A_E$
such that $y=\bigoplus_{\widehat{E}}\{ a_{\kappa} \mid \kappa\in H\}=%
\bigvee_{\widehat{E}}\{ a_{\kappa} \mid \kappa\in H\}$
(see \cite[Theorem~3.3]{ZR65}). Since
$\{ a_{\kappa} \mid \kappa\in H\}\subseteq A_E\subseteq E$
we obtain that
$\{ n_{a_{\kappa}} a_{\kappa} \mid \kappa\in H\}\subseteq S(E)$ , hence
$S(E)$ is join-dense in $S(\widehat{E})$ which gives that
$S(\widehat{E})$ is
the MacNeille completion of $S(E)$ by~\cite{schmidt}. Further
$S(\widehat{E})\cap E\subseteq S(E)$, thus
$S(\widehat{E})\cap E\subseteq S(E)\subseteq S(\widehat{E})\cap E$.

$(v)$: This follows by $(iv)$ and the
Schmidt characterization of the MacNeille completion (see~\cite{schmidt}) using the same
arguments as in the proof of $(i)$.
\end{proof}

\begin{example}
It is well known that there are
even f\/inite orthomodular lattices admitting no states (see \cite{greechie}). On
the other hand, every orthomodular lattice $(L;\vee,\wedge,{}^{\bot}, 0, 1)$
or  a Boolean algebra can be organized  into a
lattice ef\/fect algebra if we def\/ine a partial
binary operation $\oplus$ on $L$ by: $x\oplus y=x\vee y$ for every
pair $x,y\in L$ such that $x\le y^{\perp}$. This is the original
idea of G.~Boole, who supposed that $x+y$ denotes the logical
disjunction of $x$ and $y$ when the logical conjunction $xy=0$,
(see \cite{boole}). For this lattice ef\/fect algebra $(L;\oplus, 0, 1)$
the compatibility center
$B(L)=\bigcap\{ B\subseteq L\mid B \ \mbox{block of}\ L\}$ is evidently a Bolean algebra.
In spite of that a state on $L$ need not exist.
\end{example}

\begin{example}
In general, for a lattice ef\/fect algebra $E$, the compatibility center
$B(E)=\bigcap\{ B\subseteq E\mid B \ \mbox{block of}\ E\}$ is an $MV$-ef\/fect
algebra, as blocks of $E$ are $MV$-ef\/fect
algebras, (see \cite{ZR56}). Hence $B(E)$ is not a Boolean algebra if\/f
$B(E)\not\subseteq S(E)$. For instance, assume that $E_1$
is a horizontal sum of chains $\{0_1, a_1, 2a_1=1_1\}$ and
$\{0_2, a_2, 2a_2=1_2\}$ and we identify
zero and top elements. Let $E=\{0, c, 2c\}\times E_1$ be a direct product
of a chain $\{0, c, 2c\}$ and the lattice ef\/fect algebra $E_1$. Then
$E$ is a lattice ef\/fect algebra with two blocks
$M_1=\{0, c, 2c\}\times \{0_1, a_1, 2a_1\}$ and
$M_2=\{0, c, 2c\}\times \{0_2, a_2, 2a_2\}$. Hence
$B(E)=M_1\cap M_2\simeq \{0, c, 2c\}$ is not a Boolean algebra.
\end{example}

\begin{theorem}\label{teortri} Let $E$ be an atomic lattice effect algebra densely
embeddable into a complete lattice effect algebra $\widehat{E}$.
If $B(E)$ is not a Boolean algebra then there exists an
$(o)$-continuous subadditive state on~$E$.
\end{theorem}

\begin{proof} Since every complete lattice ef\/fect algebra is
 Archimedean (see \cite[Theorem 3.4]{ZR54}) so~$\widehat{E}$ is Archimedean and
the ef\/fect algebra $E$ is Archimedean, as
$E\subseteq\widehat{E}$. By \cite{mosna}
$B(E)=\bigcap \{ M\subseteq E \mid M \ \mbox{atomic} \
\mbox{block}\ \mbox{of}\ E\}\subseteq
\bigcap \{ \widehat{M}\subseteq \widehat{E} \mid
\widehat{M}, M \ \mbox{atomic} \
\mbox{block}\ \mbox{of}\ E\}=B(\widehat{E})$, by Lemma \ref{mcne}~$(ii)$. Clearly, $B(E)$ is not a Boolean algebra if\/f
there exists $x\in B(E)$ such that
$x\wedge x'\not =0$, which gives
$x\in B(\widehat{E})$ and $x\not\in S(\widehat{E})$.
Further, $C(\widehat{E})$ is atomic (Lemma \ref{mcne}~$(iii)$) and
$\bigvee_{\widehat{E}}\{p\in C(\widehat{E})
\mid p\ \mbox{atom of}\ C(\widehat{E})\}=1$,
which implies that
$\widehat{E}\cong \prod\{[0, p]\mid p\
\mbox{atom of}\ C(\widehat{E})\}$
and $B(\widehat{E})\cong \prod\{B([0, p])\mid p\
\mbox{atom of}\ C(\widehat{E})\}$. By \cite[Theorem 4.1]{PR6},
$B([0, p])=\{0, p\}$ or
$[0, p]=\{0, a, 2a, \dots, n_a a\}$ for some atom $a\in E$. Thus, there
exists an atom $p_0$ of $C(\widehat{E})$ such that
$B([0, p_0])\not= \{0, p_0\}$ and hence
$[0, p_0]=\{0, a, 2a, \dots, n_a a\}$  for some atom $a\in E$. It follows
(see also \cite[Theorem 5.8]{PR6}) that there exists
 an $(o)$-continuous subadditive state $\widehat{\omega}$ on $\widehat{E}$
and hence $\omega=\widehat{\omega}/E$ is
an $(o)$-continuous subadditive state on ${E}$ such that
$\omega(a)=\frac{1}{n_a}$.
\end{proof}

Note that conditions of Theorem  \ref{teortri} safeguard
that an $(o)$-continuous subadditive state $\omega_a$ on~$E$
exists for every atom $a\in E$ such that $a\in B(E)$ and moreover,
that at least one such atom~$a$ of~$E$ exists, as
$B(E)\not= C(E)$.

\begin{corollary}Let $E$ be an Archimedean atomic
lattice effect algebra with finitely many blocks.
If $B(E)$ is not a Boolean algebra $($equivalently
there exists $x\in B(E)$ such that
$x\wedge x'\not =0)$ then there exists an
$(o)$-continuous subadditive state on $E$.
\end{corollary}

\begin{proof} It follows from the fact that
a lattice ef\/fect algebra with f\/initely many blocks
is den\-sely
embeddable into a complete lattice ef\/fect algebra if\/f
$E$ is Archimedean (see \cite[Theo\-rem~4.5]{ZR54}).
\end{proof}

\subsection*{Acknowledgements}

This work was supported by the Slovak Research and Development
Agency under the contract No.~APVV--0375--06
and the grant VEGA-2/0032/09 of M{\v S} SR.
The author wishes to express his thanks to referees for many
stimulating questions and suggestions during the preparation of the
paper, which helped to improve it.

\pdfbookmark[1]{References}{ref}
\LastPageEnding

\end{document}